\documentstyle[12pt,epsf,cite]{article}
\begin{document}

\title{Two-Photon Interferometry for High-Resolution Imaging}
\author{Dmitry V. Strekalov and Jonathan P. Dowling.\\
Quantum Computing Technologies Group,\\
Jet Propulsion Laboratory, California Institute of Technology,\\
MS 300-123, 4800 Oak Grove Drive, Pasadena, CA 91109.\\
Dmitry.V.Strekalov@jpl.nasa.gov}
\maketitle

\begin{abstract}
We discuss the advantages of using non-classical states of light for two 
aspects of optical imaging: the creation of microscopic images on 
photosensitive substrates, which constitutes the foundation for 
optical lithography, and the imaging of microscopic objects. In 
both cases, the classical resolution limit given by the Rayleigh 
criterion is approximately half of the optical wavelength. It has 
been shown, however, that by using multi-photon quantum states of the 
light field, and a multi-photon sensitive material or detector, this 
limit can be surpassed. We give a rigorous 
quantum mechanical treatment of this problem, address some 
particularly widespread misconceptions, and discuss turning quantum imaging 
into a 
practical technology.
\end{abstract}

The idea of overcoming the limits of classical optical imaging 
by using multi-photon processes is fairly well known. For example, 
Marlan Scully discusses, in his book \cite{scullybk}, a two-photon 
microscope scheme that beats the diffraction limitation by a factor of 
$\sqrt{2}$, 
by making a 
${\rm sinc}^4(kx)$ diffraction pattern instead of the usual ${\rm sinc}^2(kx)$. 
Such 
narrowing of a diffraction pattern can be observed by a 
detector sensitive to the square of intensity, instead of just intensity 
itself. In 
other words, one needs a two-photon process to observe the 
$\sqrt{2}$ narrowing beyond the diffraction limit, even within classical 
optics. Moreover, using 
detectors based on a higher-order multi-photon process, which are 
sensitive exclusively to the higher orders of intensity, one 
could see even narrower diffraction patterns. 

This approach would not work so well for holographic imaging used in 
lithography. In this technique, the desired image is composed of 
interference fringes of different spatial frequencies, so the 
resolution is given by the highest spatial frequency. This spatial 
frequency is equal to the inverse of the fringe period, which cannot 
be shorter than one half of the optical wavelength. It is easy to see 
that this period is the same for any power of intensity, e.g. a 
$\sin^4(kx)$ fringe has the same period as a $\sin^2(kx)$ fringe. 

Different approaches have been suggested to obtain an  
interference fringe of the square of the intensity with a shorter period. 
It has been proposed, for example, that frequency modulation can be used to 
blur the longer-wavelength component of a $\sin^4(kx)$ fringe 
\cite{yablonovitch99}. 
The use of 
quantum sources of light to beat this limit has also been proposed 
\cite{boto00} and 
demonstrated with electronic coincidence detection \cite{dangelo01}. 

\begin{figure}[htp]
\centerline{
\input epsf
\setlength{\epsfxsize}{3.3in}
\setlength{\epsfysize}{1.9in}
\epsffile{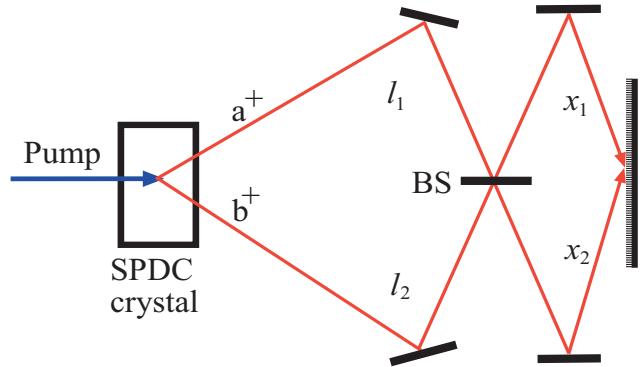}
}
\caption[setup]{\label{fig:setup}Two-photon interferometer with 
photosensitive substrate.}
\end{figure}

Consider the setup in Fig.\ref{fig:setup} that has been proposed for quantum 
interferometric 
lithography \cite{boto00}. This is a 
modification of a well-known two-photon interference experiment 
\cite{hong87,shih88}, in which the single-photon detectors are 
removed and the output beams are directed at a two-photon sensitive 
substrate (e.g., one covered with a lithographical photoresist).

Following the standard theoretical treatment for two-photon 
interferometers, we write the two-photon square amplitude as
\begin{equation}
 |A|^2 \equiv \langle \Psi| \hat{E}^{(-)} \hat{E}^{(-)} \hat{E}^{(+)} 
\hat{E}^{(+)}|\Psi\rangle = |\langle 0| \hat{E}^{(+)} 
\hat{E}^{(+)}|\Psi\rangle|^2 ,
 	\label{A}
\end{equation}
where the fields depend on the propagation paths, and the state 
$|\Psi\rangle$ is the frequency-entangled output state of a 
Spontaneous Parametric Down Converter (SPDC):
\begin{equation}
|\Psi\rangle  = \int d\nu h(\nu) \hat{a}^\dagger(\nu) 
\hat{b}^\dagger(-\nu) |0\rangle.
\label{Psi}
\end{equation}
In (\ref{Psi}), creation operators $ \hat{a}^\dagger $ and $ 
\hat{b}^\dagger$ refer to channels labeled $l_1$ and $l_2$, 
respectively, in Fig.\ref{fig:setup}; $\nu$ is the frequency-detuning 
from the central frequency $\omega_0$, the later being equal to one 
half of the pump frequency $\omega_p$. The spectral function $h(\nu)$ 
gives the phase-matching width and accounts for inexact momentum 
conservation due to the finite length $L$ of the crystal:
\begin{equation}
h(\nu) = \frac{1-e^{-iL\Delta_z(\nu)}}{ iL\Delta_z(\nu)}.
\label{h}
\end{equation}
Derivation and analysis of expressions (\ref{Psi}) and (\ref{h}) are 
given in a number of publications on SPDC. In particular, in 
\cite{shih94c,burlakov97}, it is shown that for collinear degenerate 
type-I SPDC
\begin{equation}
\Delta_z(\nu) = -D'\nu^2, \quad D'=\frac{d}{d\omega}\frac{1}{v}
\stackrel{\mid}{\mid}_{\omega_0},
\label{delta1}
\end{equation}
and for collinear degenerate type-II SPDC, where the signal and idler photons 
have orthogonal polarizations, we have
\begin{equation}
\Delta_z(\nu) = D\nu, \quad D=\frac{1}{v_o}-\frac{1}{v_e},
\label{delta2}
\end{equation}
where $v$ denotes the group velocity of the signal and idler photons. 
In case of orthogonal polarizations (type-II), the group velocity $v$ 
has indices $o$ and $e$ for ``ordinary" and ``extraordinary" 
polarization components.

The two-photon amplitude of Eq.(\ref{A}) can describe the coincidence detection 
rate, as well as the two-photon absorption rate, as a function of 
pathlengths $l_{1,2}$ and $x_{1,2}$. In the coincidence detection 
case, the fields in Eq. (\ref{A}) are evaluated at the two distinct locations 
of the two detectors, while in the two-photon absorption case they are 
evaluated at the same, arbitrary, point on the 
photosensitive substrate. A geometric size of the ``point" in this 
context may be equal to the size of photo emulsion grain, or of the 
photoresist molecule. It is reasonable to assume that this size is much smaller 
than the interference structure we are expecting to see. As a further 
simplification, we 
will consider a one-dimensional problem with exactly 
counterpropagating beams. This geometry is obviously not practical, 
since no light energy is delivered to the surface, and we study this 
case just as an illustration allowing us to simplify the treatment. 

As a next step, we need to represent the fields in (\ref{A}) in terms 
of the same operators that describe the two-photon wavefunction, Eq. 
(\ref{Psi}). For perfectly monochromatic plane waves with a wavenumber 
$k=\omega_0/c$, the representation is obtained by propagating the 
operators through the interferometer:
\begin{equation}
\hat{E}^{(+)}= \hat{a} e^{ikl_1}\left (\frac{1}{\sqrt{2}} 
e^{ikx_2}+\frac{i}{\sqrt{2}} e^{ikx_1}\right)+
\hat{b} e^{ikl_2}\left (\frac{1}{\sqrt{2}} 
e^{ikx_1}+\frac{i}{\sqrt{2}} e^{ikx_2}\right).
\label{E0}
\end{equation}
In equation (\ref{E0}), we put the proportionality constant between 
the field operator and the annihilation operator equal to unity. 
Also, we assume that the fields in the arms $l_1$ and $l_2$ have the 
same polarization. It is easy to see that otherwise there will be no 
two-photon interference fringes on the photosensitive substrate.

\begin{figure}[htp]
\centerline{
\input epsf
\setlength{\epsfxsize}{4.5in}
\setlength{\epsfysize}{3.6in}
\epsffile{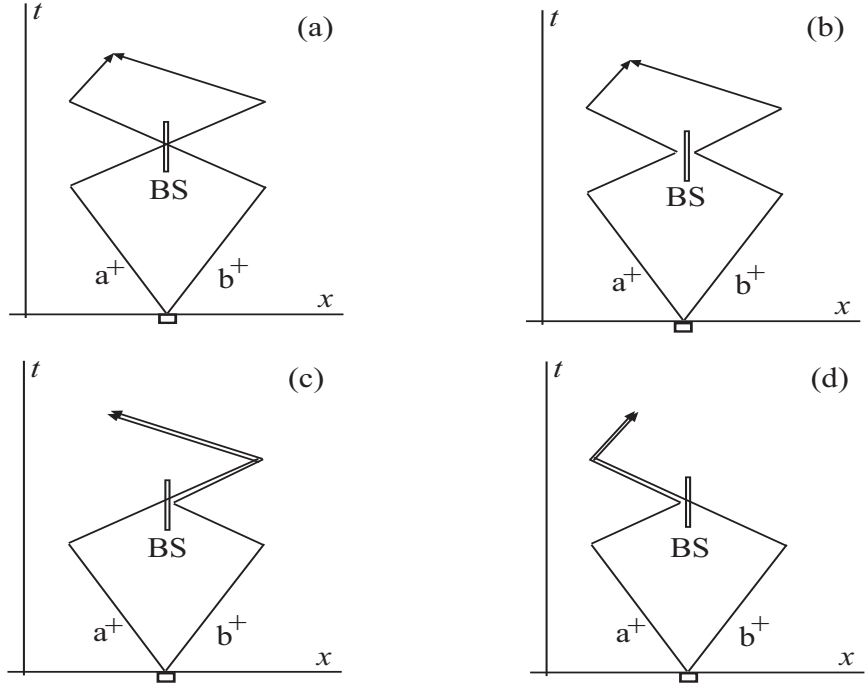}
}
\caption[paths]{\label{fig:paths}Different two-photon paths 
contributing to the amplitude (\ref{A}): (a) both photons are 
transmitted; (b) both reflected; (c) transmitted - reflected; (d) 
reflected - transmitted.}
\end{figure}

The plane-wave approximation implies that in the wave function Eq. 
(\ref{Psi}), $h(\nu)$ should be replaced by $\delta(\nu)$. Then 
substituting Eqs. (\ref{E0}) and (\ref{Psi}) into Eq. (\ref{A}) it is easy to 
notice that the terms with $\hat{a}^2$ and $\hat{b}^2$ drop out, 
which is consistent with only one photon being present in each channel. The 
other four terms can be represented by four paths shown in 
Fig.\ref{fig:paths}. These paths correspond to both photons being 
transmitted by the beamsplitter (a), both reflected by it (b), one 
transmitted, the other reflected (c), and vice versa (d).
Notice that, in the usual coincidence-detection treatment of 
two-photon interference, the amplitudes corresponding to paths (c) 
and (d) are discarded simply because they do not result in a pair of 
coincident detections. Therefore one cannot directly apply to our system the 
results well known for a two-detectors experiment, and 
then argue that the detectors are placed at the same point, 
since this leads to loss of the amplitudes (c) and (d). Let us 
now show that it is \emph{these} amplitudes that give rise to two-photon 
interference.

In the following, we will consider the more realistic model of 
wavepackets rather than plane waves. The fields will be allowed to 
have a finite frequency bandwidth around the central frequency $\omega_0$, 
described by a real, even function $f(\nu)$:

\begin{equation}
E  =  \int d\nu f(\nu) \label{E}
\left\{ \hat{a}(\nu)e^{ik(\nu)l_1} \left(e^{ik(\nu)x_2}+ i 
e^{ik(\nu)x_1}\right)+ \hat{b}(\nu) e^{ik(\nu)l_2} 
\left(e^{ik(\nu)x_1}+ ie^{ik(\nu)x_2}\right)\right\}.
\end{equation}
Then the two-photon amplitude, Eq. (\ref{A}), takes on the following form:
\begin{eqnarray}
A  &= &\int d\nu d\nu_1 d\nu_2 h(\nu) f(\nu_1) f(\nu_2) \nonumber \\
&&\left\{ e^{ik(\nu_1)l_1} e^{ik(\nu_2)l_2} \left(e^{ik(\nu_1)x_2}+ i 
e^{ik(\nu_1)x_1}\right) \left(e^{ik(\nu_2)x_1}+ i 
e^{ik(\nu_2)x_2}\right)\right\}\label{A1}\\
&&\langle 0| \hat{a}(\nu_1) \hat{b}(\nu_2)  \hat{a}^\dagger(\nu) 
\hat{b}^\dagger(-\nu)|0\rangle .
\nonumber
\end{eqnarray}
The inner product in Eq. (\ref{A1}) is equal to 
$\delta(\nu_1-\nu)\delta(\nu_2+\nu)$ which reduces Eq. (\ref{A1}) to a 
single integral. To handle it, we expand $k(\nu)=k_0+\nu/c$, where 
$k_0\equiv k(\omega_0)$. This allows us to arrive at,
\begin{equation}
A  =  e^{ik _0(l+x)} \left [u(\Delta l+\Delta x)- u(\Delta l-\Delta 
x)+2u(\Delta l) \cos(2k_0\Delta x) \right],
\label{Au}
\end{equation}
where $u(z)$ is given by a Fourier transformation of a combined 
spectral density, and therefore has a meaning of a correlation 
function, namely,
\begin{equation}
u(z)\equiv \int d\nu h(\nu) f^2(\nu) e^{i\frac{\nu}{c}z}.
\label{u}
\end{equation}
In Eq. (\ref{u}), the variables $x\equiv x_2+x_1$, $\Delta x\equiv 
\frac{1}{2}(x_2-x_1)$, $l\equiv 
l_2+l_1$ and $\Delta l\equiv l_2-l_1$ have been introduced. Note that 
coordinate along the substrate $\Delta x$ is equal to a half of the path 
difference $x_2-x_1$.

Analyzing the symmetry properties of $h(\nu)$, we find that in both 
cases of type-I, Eq. (\ref{delta1}), and type-II, Eq. (\ref{delta2}), SPDC 
$u(z)$ 
is always a real, even function:
\begin{equation}
u(z) = u(-z)= u^*(z)= u^*(-z).
\label{ueven}
\end{equation}
Therefore the first two terms in Eq. (\ref{Au}) cancel each other when 
$\Delta l =0$. Taking the absolute square of the remaining term, we 
get
\begin{equation}
|A|^2 = 4u^2(0) \cos^2(2k_0\Delta x).
\label{finalA}
\end{equation}
We see from Eq. (\ref{finalA}) that the two-photon absorption amplitude 
is a perioic function of coordinate $\Delta x$ measured along the 
photosensitive substrate, that has a spatial frequency $4k_0$, which 
is twice the spatial frequency of the usual, second-order 
interference fringes. The two-photon interference fringes of Eq. 
(\ref{finalA}) appear to have a perfect contrast for all $\Delta x$. 
This is a consequence of a plane-wave approximation {\it for the 
pump}. If one considers the pump with a finite bandwidth, the 
exponential pre-factor in (\ref{Au}) will no longer be just a phase 
factor, but will turn into an envelope, equivalent to the pump 
envelope. Therefore the two-photon interference fringes 
(\ref{finalA}) will have a coherence length equal to the pump 
coherence length, which may be quite long and can reach meters for cw 
lasers. 

It is very important that the two-photon coherence length does not 
depend on the bandwidth of the fields given by $f(\nu)$, nor on the 
phase-matching width given by $h(\nu)$. This is obvious from 
the condition $\Delta l =0$. It has been 
shown \cite{hong87,shih88}, that in this case the two-photon 
amplitudes represented in Fig.\ref{fig:paths} by diagrams (a) and (b) 
exactly cancel each other, and the photon pair always goes to one 
channel (either $x_1$ or $x_2$), depicted in diagrams (c) or (d). In 
other words \cite{boto00}, the beamsplitter produces an entangled state, 
$|2\rangle_{x_1}|0\rangle_{x_2}-|0\rangle_{x_1}|2\rangle_{x_2}$, 
which picks up spatial phase at the same rate as the pump 
photon would. It also dephases at the same slow rate as the pump 
photon does, due to its finite bandwidth, which results in the 
two-photon coherence length of the SPDC light being equal to the pump 
(single-photon) coherence length.

Now let us consider the linear interference in our apparatus. This 
analysis is important, since the modulations of intensity will directly 
affect the result Eq. (\ref{finalA}) for the two-photon absorption rate. For 
example, there will be no two-photon absorption in the nodes of the 
single-photon interference fringe. 

The expression for intensity is
\begin{equation}
I = \langle \Psi|\hat{E}^{(-)} \hat{E}^{(+)}|\Psi\rangle,
\label{I}
\end{equation}
where the state $|\Psi\rangle$ is given by Eq. (\ref{Psi}) and the field 
is given by Eq. (\ref{E}). Setting $l_1 = l_2$, and treating this 
expression the same way we have treated the forth-order field 
momenta, we arrive at
\begin{equation}
I = 1- \cos(2k_0\Delta x)\int d\nu |h(\nu)|^2 f^2(\nu) \sin\left 
(2\frac{\nu}{c}\Delta x\right).
\label{finalI}
\end{equation}
Notice that the integrand in Eq. (\ref{finalI}) is an odd function, and 
hence the whole integral is zero and Eq. (\ref{finalI}) 
equals unity. This means that in our apparatus there will be no 
intensity modulations due to the second-order interference, 
regardless of individual coherence length of the signal and idler 
photons. This at first appears surprising, since one might expect to 
see at least a few interference fringes at the white light 
interference condition $x_1 = x_2$. However, taking into account 
that both inputs of the beamsplitter are used, we realize that we 
actually have two sets of interference fringes exactly out of phase 
with each other, and hence the total intensity is unmodulated. 

Two more issues associated with two-photon quantum imaging need to be 
addressed to make it a practically useful technology. One is the 
availability of two-photon sensitive photoresists and detectors, and 
the other has to do with the fact that using SPDC as a two-photon 
source, one first loses a factor of two in spatial resolution by down 
converting the pump frequency (and hence doubling the wavelength), 
and then re-gains this factor by using two-photon processes. 
Therefore, in terms of spatial resolution, our quantum imaging technique 
has no apparent advantage over using classical imaging at the pump wavelength. 
The counter is that it is not always 
possible to use the UV light argument \cite{belfield00}. For example, it may be 
incompatible 
with imaging biological or other light sensitive objects. Another 
example is 3D lithography \cite{belfield00}. Creating 3D structures 
with single-photon exposure of photolithographical materials is very 
difficult, since they strongly absorb UV light, which limits the depth 
of penetration. Two-photon exposures solve this problem. However, 
much value would be added to the quantum imaging technology if one 
could prepare two-photon states without doubling the wavelength. One 
way to achieve it is  to use a Hyper Parametric Scattering (HPS) 
instead of SPDC.

\begin{figure}[htp]
\centerline{
\input epsf
\setlength{\epsfxsize}{5in}
\setlength{\epsfysize}{2in}
\epsffile{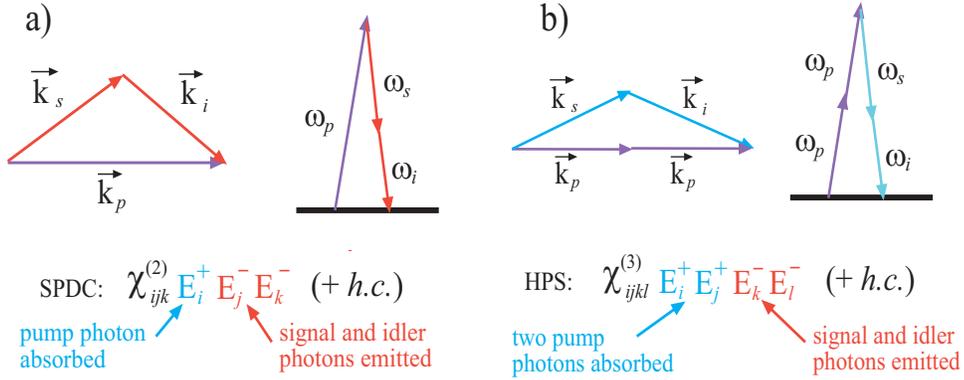}
}
\caption[matching]{\label{fig:matching}The phase-matching (momentum 
and energy conservation) diagrams for SPDC (a) and HPS (b).}
\end{figure}

HPS is a nonlinear optical process occurring via the cubical optical 
nonlinearity $\chi^{(3)}$, in which {\it two} pump photons recombine 
into an entangled photon pair. This process is similar to four wave 
mixing in the same sense as SPDC is similar to Parametric 
Amplification: four wave mixing and PA assume non-vacuum input into 
the signal or the idler modes. HPS is distinct from the SPDC, 
where {\it a single} pump photon produces an entangled pair. This 
distinction is most evident from comparing the phase-matching 
conditions for SPDC with those for HPS:
\begin{eqnarray}
\vec{k}_p = \vec{k}_s +\vec{k}_i, &\quad& \omega_p = \omega_s + \omega_i,\\
\label{spdcmatch}
2\vec{k}_p = \vec{k}_s +\vec{k}_i, &\quad& 2\omega_p = \omega_s + 
\omega_i,
\label{hpsmatch}
\end{eqnarray}
which is illustrated graphically in Fig.\ref{fig:matching}. An 
important thing to notice in Fig.\ref{fig:matching} is that the 
average wavelength of the photons produced in HPS is the same as that 
of the pump, while in the case of SPDC it doubles.

HPS was observed for the first time over 30 years ago 
\cite{weinberg69}. At that time, it did not attract attention 
as a source of EPR-states because of a very low efficiency of the 
$\chi^{(3)}$ processes compared to $\chi^{(2)}$ processes. A typical 
value for $\chi^{(2)}$ is $10^{-8}$ [CGS units of electric 
field]$^{-1}$, while for $\chi^{(3)}$ it is $10^{-15}$ [CGS units of 
field]$^{-2}$. Fortunately, HPS output power is quadratic with 
respect to the pump intensity, while in case of SPDC it is only 
linear. To compare efficiencies of the two processes, one must compare 
the $\mid E_p \chi^{(2)}\mid$ and $\mid \chi^{(3)}\mid$. Modern powerful 
femtosecond lasers, that were not yet available in the early days of 
HPS, dramatically changed the situation in favor of HPS.

Another argument in favor of HPS is that, unlike SPDC, this process 
does not require any particular symmetry of the media and can be 
observed not only in crystals but also in glass fibers 
\cite{kumar94}, which promises to increase the interaction length to 
meters or beyond. Furthermore, it has been shown \cite{klyshkobk} 
that nearly four orders of magnitude improvement of the signal can 
be achieved by cascading two $\chi^{(2)}$ processes to emulate a 
$\chi^{(3)}$ HPS process. A large amount of research has been 
done on $\chi^{(3)}$ processes, and particularly on four wave mixing 
\cite{slusher85,kumar94,nagasako97,boyd99}, and we plan to rely 
on these results in our new research program directed at creating a 
robust source of entangled photon pairs or two-photon states without 
down converting from a higher frequency.

The second practical issue, mentioned above, is the availability of 
two-photon sensitive photoresists. Considering the very low power of 
two-photon sources, a high two-photon sensitivity of the photoresists 
is required. Unfortunately, high, single-photon, UV sensitivity of many 
commercially available photoresists does not guarantee that they would 
be suitable two-photon sensitive materials. The synthesis of such a 
material appears to be a difficult task, although a large volume of 
research has been done in this area motivated by the growing 
recognition of the two-photon imaging technology importance 
\cite{borisov98a,joshi99,belfield00}.

We also have carried out a preliminary search for two-photon 
sensitive lithographic materials. Relying on the analogy with atomic 
systems, we expect that a suitable two-photon material would have an 
intermediate level corresponding to the single photon energy, so that 
the single-photon detuning, which factors inversely into the two-photon 
absorption cross section, is small, and the two-photon absorption rate 
is peaked. It is furthermore required that the molecular transition 
corresponding to the intermediate absorption level does not result in 
the photochemical reaction in the phototresist (otherwise the 
resist would be one-photon sensitive). Finally, we require that the 
intermediate level or 
band is normally depopulated and very short-lived (otherwise the 
resist would be one-photon sensitive via cascaded processes); and 
that both transitions have the correct selection rules.  

\begin{figure}[htp]
\centerline{
\input epsf
\setlength{\epsfxsize}{4.5in}
\setlength{\epsfysize}{3in}
\epsffile{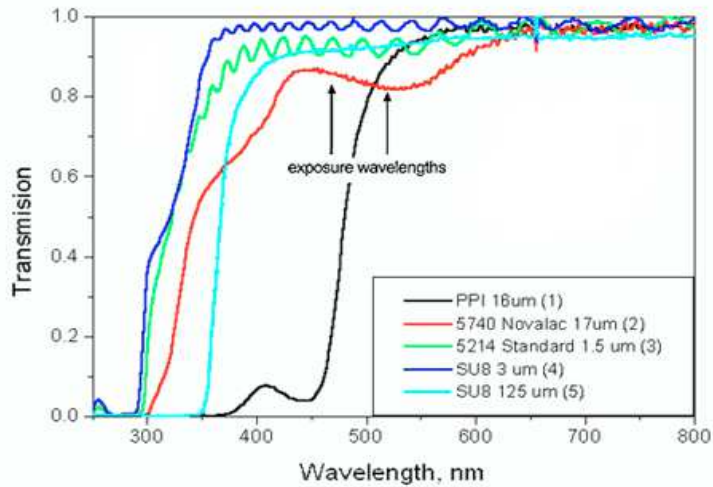}
}
\vspace*{-0.3in}
\caption[spectra]{\label{fig:spectra} Absorption spectra for 
different photoresists. 
The sample of the choice shows an absorption maximum centered at 
about 520 nm. 
Arrows mark the wavelengths the sample was exposed at: 457.9 nm and 
514.5 nm.}
\end{figure}

We have taken absorption spectra of various commercially available 
photoresists. The results are shown in Fig.\ref{fig:spectra}. One of 
our samples, the Novalac 5740, has shown a local absorption maximum 
which is centered at about 520 nm and is clearly separated from the 
strong transition in the UV part of the spectra, which is associated with the 
photochemical reaction initiating the photoresist. We spun an 
approximately 15 $\mu$m-thick sample of this photoresist on a gold 
plated substrate and exposed the sample to different doses of the 
Argon Ion laser light, whose wavelength (514.5 nm) was close to the 
center of the absorption peak of interest. We found the threshold 
dose of about 2 kJ/cm$^2$, assuming 100\% radiation reflection off 
the mirror substrate and operating at the intensity level of 5 
W/cm$^2$. Repeating the experiment at 25 W/cm$^2$, we obtained the 
same results with an exposure time that was five times shorter. This result 
suggests 
that the exposure process is linear in intensity and hence is a 
single-photon one. Notice that the threshold we found at 514.5 nm is 
roughly five orders of magnitude higher then for a regular UV 
exposure.

Next, we repeated the exposures for another Argon Ion laser line 
with the 457.9 nm wavelength, which is off the intermediate 
absorption peak but is closer to the UV absorption transition. We 
found that at this wavelength the threshold dose was definitely lower 
than 0.4 kJ/cm$^2$. This suggests that the high-threshold 
photo-initiation observed at 514.5 nm, as well as at 457.9 nm, is not 
related to the intermediate absorption peak, but rather is due to a 
far off-resonant absorption on the wing of the UV absorbing 
transition. Therefore, the selected material may satisfy the 
above-outlined requirements for a two-photon optimized photoresist, 
and it would be interesting to try exposing it with a two-photon 
source.  We plan on carrying out such experiment in the nearest 
future.

In conclusion, we have carried out a rigorous analysis that confirmed 
the earlier results \cite{boto00}. In addition, our analysis has shown that 
the desired two-photon interference fringe will have a very long 
coherence length, equal to that of the pump, and that the 
second-order (single-photon) interference fringes will be entirely 
absent. The questions related to alternative sources of two-photon 
states and to the choice of two-photon sensitive photo-lithographical 
materials have been discussed. Although bringing the research in this 
area to the level of practical technology is a challenging task, it 
is at the same time is an interesting and potentially rewarding one.

{\bf Acknowledgments }

We would like to acknowledge useful discussions with R. Y. Chiao, J. D. 
Franson, 
D. F. V. James P. G. Kwiat,  W.D. Phillips, Y. H. Shih, J. E. Sipe, and 
A. M. Steinberg. We thank Victor White for valuable help with photoresists.
We would like to acknowledge financial support from the National Aeronautics 
and Space Administration, the Office of Naval Research, and the 
Advanced Research and Development Activity.

\bibliography{titles}
\end{document}